\documentclass[preprint,5p]{elsarticle}

\biboptions{sort,compress}

\usepackage[scaled]{helvet}
 
\usepackage[T1]{fontenc}

\usepackage{nomencl}
\makenomenclature
\renewcommand*{\nompreamble}{\footnotesize} % Or \small, \scriptsize
\renewcommand*\nompreamble{\begin{multicols}{2}}
\renewcommand*\nompostamble{\end{multicols}}

\usepackage{booktabs} % For better horizontal lines (\toprule, \midrule, \bottomrule)
\usepackage{enumitem} % For customizing itemize lists\
\usepackage{array}

\usepackage{amssymb}
\usepackage{amsmath}
 \usepackage{lineno}

\usepackage{hyperref}

\usepackage{cleveref}
\crefformat{figure}{{Fig.}~#2#1#3}
\Crefformat{figure}{{Fig.}~#2#1#3}
\crefformat{table}{{Tab.}~#2#1#3}
\Crefformat{table}{{Tab.}~#2#1#3}

\crefformat{section}{{Sect.}~#2#1#3}
\Crefformat{section}{{Sect.}~#2#1#3}

\crefformat{subsection}{{Sect.}~#2#1#3}
\Crefformat{subsection}{{Sect.}~#2#1#3}

\begin{document}

\begin{frontmatter}

\title{Mechanistic inference of stochastic gene expression from structured single-cell data\tnoteref{t1}} %% Article title

\tnotetext[t1]{Submitted invited review for the  `Identifiability, estimation and uncertainty in mathematical modelling' issue in \textit{Current Opinion in Systems Biology}.}

\author[uci,utah]{Christopher E. Miles}

\ead{chris.miles@uci.edu}

 \affiliation[uci]{organization={Department of Mathematics, Center for Complex Biological Systems, University of California, Irvine},
 %            addressline={},
 %            city={},
%             postcode={},
             ,state={CA}
             ,country={USA}}

 \affiliation[utah]{organization={Present address: Department of Mathematics, University of Utah},
          %  addressline={},
          %   city={},
           %  postcode={},
             ,state={UT}
             ,country={USA}}

%% Abstract
\begin{abstract}
Single-cell gene expression measurements encode variability spanning molecular noise, cell-to-cell heterogeneity, and technical artifacts. Mechanistic stochastic models provide powerful approaches to disentangle these sources, yet inferring underlying dynamics from standard snapshot sequencing data faces fundamental identifiability limitations. This review focuses on how structured datasets with temporal, spatial, or multimodal features offer constraints to resolve these ambiguities, but demand more sophisticated models and inference strategies, including machine-learning techniques with inherent tradeoffs. We highlight recent progress in the judicious integration of structured single-cell data, stochastic model development, and innovative inference strategies to extract predictive, gene-level insights. These advances lay the foundation for scaling mechanistic inference upward to regulatory networks and multicellular tissues.\end{abstract}

%%Research highlights
%\begin{highlights}
%\item Mechanistic models reveal interpretable structure in expression variability
%\item Temporal, spatial, and multimodal data reshape what can be inferred
%\item Increased data and model complexity challenge identifiability and reliability
%\item Simulation-based and neural methods extend inference to complex models, but require care to ensure reliability
%\item Principled single-gene inference builds foundations for understanding complex regulatory systems
%\end{highlights}

%% Keywords

\end{frontmatter}

\vspace{6pt} % Optional: add some space
\noindent\textbf{Highlights}
\begin{itemize}
\item Mechanistic models reveal interpretable structure in expression variability
\item Temporal, spatial, and multimodal data reshape what mechanisms can be inferred
\item Increased data and model complexity challenge identifiability and reliability
\item Simulation-based and ML methods expand inference, but require careful validation
\item Principled single-gene approaches lay foundations for regulatory \& multicellular systems
\end{itemize}

\section{Introduction}

\begin{figure*}[htb]
    \centering
    \includegraphics[width=0.95\linewidth]{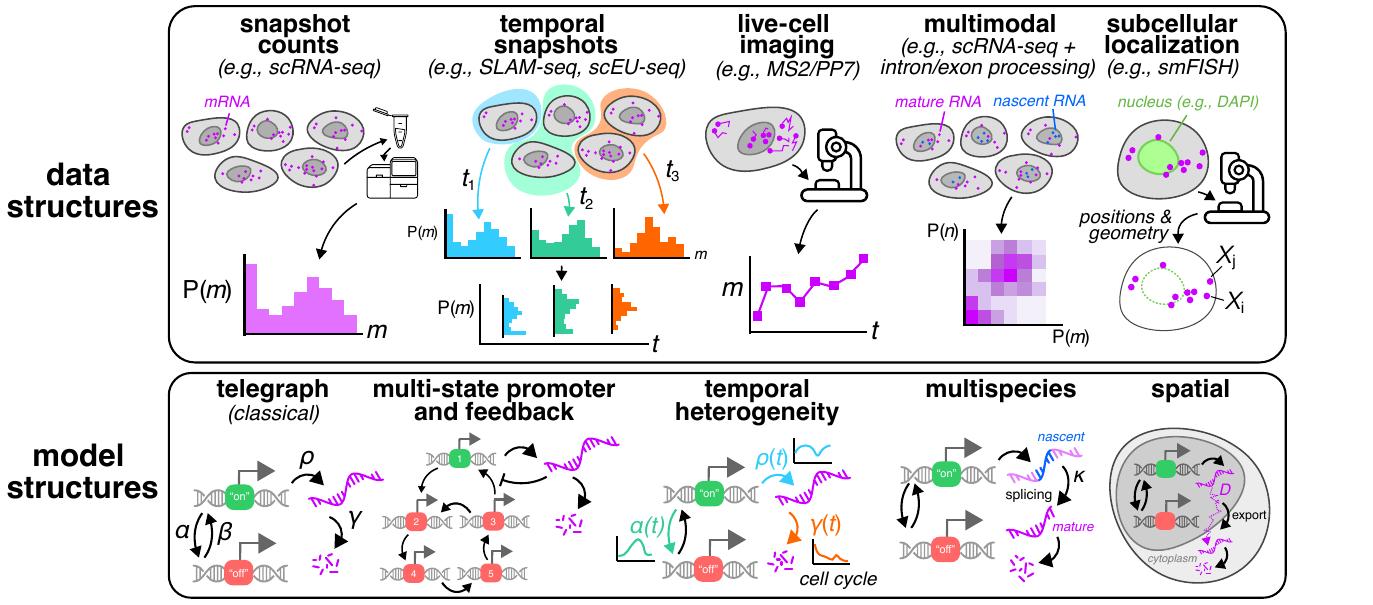}
    \caption{Landscape of single-cell gene expression data and model structures considered throughout the review. Data with additional structure can resolve model ambiguities, but necessitates developing new models and inference techniques.}
    \label{fig:data-models}
\end{figure*}

%\linenumbers

Single-cell gene expression measurements encode a tapestry of variability  spanning intrinsic molecular noise, cellular heterogeneity, and confounding technical artifacts. Far from mere nuisance, the biological portion of these fluctuations shapes cellular identities~\cite{coomer2022NoiseDistortsEpigenetic} and disease progression~\cite{alachkar2023VariabilityInnateImmune}. 
Disentangling the interwoven noise components is therefore essential for extracting causal understanding of genetic processes from single-cell data. Beyond statistical pattern-finding, mechanistic stochastic models offer an explanatory framework for this task by linking specific molecular and experimental processes to observed distributions~\cite{gorin2023StudyingStochasticSystems}. Application of such models to single-cell data has empowered numerous quantitative insights, from cataloging transcriptional dynamics across the human genome~\cite{dar2012TranscriptionalBurstFrequency} to understanding how these kinetics are regulated~\cite{larsson2019GenomicEncodingTranscriptional,luo2023GenomewideInferenceReveals,mahat2024SinglecellNascentRNA,chen2025ConservedCouplingTranscriptional} and vary by cell state~\cite{gao2024DissectionIntegrationBursty,chari2024BiophysicallyInterpretableInference}. More than just descriptive, these models provide mechanistic parameters that move beyond observing \textit{whether} a system has changed to explaining \textit{how} its underlying molecular processes are altered under perturbation, a distinction essential for therapeutic design and cellular engineering.

Despite successes in dissecting transcriptional dynamics, reliably inferring kinetics and distinguishing between hypotheses using typical high-throughput single-cell data remains challenging. Standard `snapshot' measurements of transcript counts from conventional scRNA-seq data often lack sufficient information to uniquely constrain parameters or discriminate between models. This review examines how structured single-cell datasets, sophisticated stochastic models, and advanced inference strategies, together, can overcome these limitations to disentangle the dynamics.
We focus on inference for individual genes, which, despite their apparent simplicity, pose surprising inferential obstacles and provide a valuable foundation for understanding regulatory and multicellular scales.
We first illustrate fundamental challenges in mechanistic inference from snapshot data, then explore how richer experimental designs incorporating temporal dynamics (e.g., metabolic labeling), spatial organization, and multimodal measurements (e.g., nascent/mature RNA) provide critical constraints for interpretation (see \cref{fig:data-models}). As datasets and models grow in complexity, we evaluate the evolving toolkit of inference methods, from classical approaches to machine learning techniques, with particular attention to reliability challenges introduced by approximations and surrogate models.  Throughout, we emphasize that the integration of appropriate data structures, models, and carefully validated inference is essential for extracting reliable biological insights from single-cell data.

\section{Promises and Challenges of Mechanistic Inference with the Telegraph Model}

\label{sect:teleg}

A natural entry point into model-based inference with single-cell data is the classical telegraph model. This model minimally captures transcriptional bursting (the stochastic switching between transcriptionally active and silent states \cite{tunnacliffe2020WhatTranscriptionalBurst}) and has thus served as a cornerstone of quantitative single-cell analyses. By examining the model's promises and limitations, we identify broader challenges in mechanistic inference that motivate the need for richer data and more flexible models.

The telegraph model describes a gene promoter stochastically switching between inactive $G$ (``off'') and active $G^*$ (``on'') states with respective rates $\alpha$ and $\beta$. While in the active state, transcription produces mRNA at rate $\rho$, and mRNA is degraded at rate $\gamma$. These processes are summarized by the scheme
$$
G \overset{\alpha}{\underset{\beta}{\rightleftarrows}} G^*, \quad G^* \overset{\rho}{\rightarrow} G^*+ m, \quad m \overset{\gamma}{\rightarrow} \varnothing.
$$
 Observations of the model are encoded by the \textit{chemical master equations} for $P_G(m,t)$  and $P_{G^*}(m,t)$, the probability distributions of mRNA counts in each promoter state with $m=0,1,\ldots$,
{\scriptsize
\begin{align}
\mathrm{d} P_{G} / \mathrm{d} t &= \beta P_{G^*}(m,t) + \gamma (m+1) P_{G}(m+1, t) \nonumber \\
&\quad - (\alpha + \gamma m) P_{G}(m,t) \label{eq:cme_G_prev} \\
\mathrm{d} P_{G^*}/ \mathrm{d} t &= \alpha P_{G}(m,t) + \rho P_{G^*}(m-1, t) \nonumber \\&\qquad + \gamma (m+1) P_{G^*}(m+1, t)  - (\beta + \rho + \gamma m) P_{G^*}(m,t) \label{eq:cme_Gstar_prev}
\end{align}}

The resulting steady-state distribution for $P(m) = P_G(m)+P_{G^*}(m)$  is the key to both the model's utility and inferential challenges. It classically follows a Poisson-Beta distribution with three effective parameters: the relative promoter switching rates $a=\alpha/\gamma$ and $b=\beta/\gamma$, and relative synthesis rate $r=\rho/\gamma$. This Poisson-Beta distribution can reproduce the unimodal or bimodal observations of mRNA counts, and reduces to the widely observed negative binomial distribution in the limit of short, infrequent pulses of transcription ($\alpha \ll \gamma, \alpha \ll \beta$) \cite{kim2013InferringKineticsStochastic}. This analytical simplicity provides a valuable basis for extensions that
address practical experimental complexities. For example, modeling measurement
noise (e.g., mRNA capture efficiency) upon the telegraph framework improves
inference quality \cite{gorin2022DistinguishingBiophysicalStochasticity,
tang2023ModellingCaptureEfficiency,grima2024QuantifyingCorrectingBias}.

\begin{figure}[htb]
    \centering
    \includegraphics[width=\linewidth]{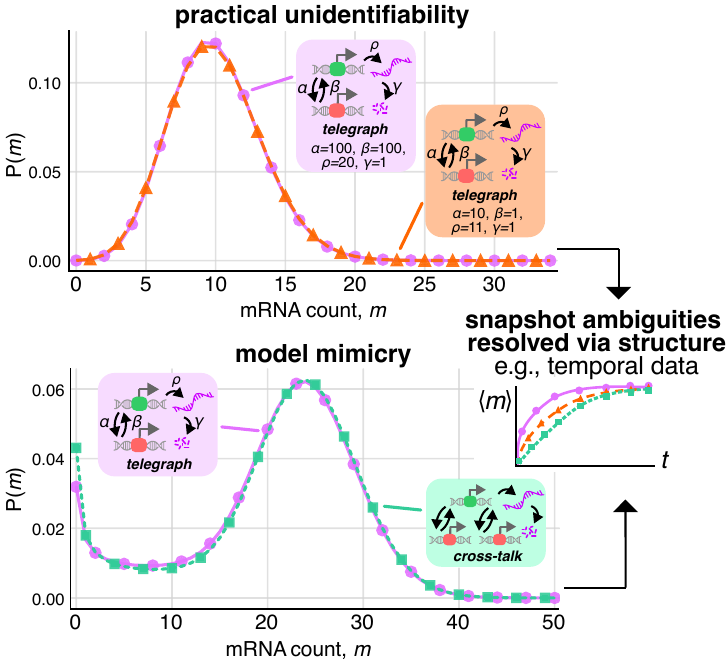}
    \caption{Inference challenges with snapshot observations of the telegraph model. \textit{Practical identifiability:} Starkly different kinetics can produce the same distribution, shown in \cite{guScalableInferenceIdentifiability}. \textit{Model mimicry:} More complex promoter structures, such as a crosstalk pathway model, can produce a nearly identical distribution, shown in \cite{jiao2024WhatCanWe}.}
    \label{fig:telegraph-issues}
\end{figure}

However, the resulting Poisson-Beta distribution highlights a fundamental inference challenge: the original four kinetic parameters in the model cannot be resolved from steady-state snapshots alone. Such a limitation is an example of \textit{structural unidentifiability}, where model parameters (or combinations thereof) can fundamentally not be resolved from particular data \cite{cinquemani2018IdentifiabilityReconstructionBiochemical}. Even these three effective parameters are challenging to infer reliably. The three-parameter distribution exhibits \textit{practical unidentifiability}, where different parameter combinations yield nearly identical distributions (\cref{fig:telegraph-issues}). This parameter ambiguity is exacerbated by technical noise \cite{guScalableInferenceIdentifiability,wang2025NoiseModelsNumbers} and by extrinsic variability like cell-cycle progression \cite{ham2021PathwayDynamicsCan,fu2022QuantifyingHowPosttranscriptional}. 
One might assume these inferential obstacles stem from the model's simplicity. However, fundamentally more complex variants (e.g., involving multiple promoter states or feedback) can generate snapshot distributions statistically indistinguishable from the simple telegraph model, a phenomenon called \textit{model mimicry} in \citet{jiao2024WhatCanWe}, shown in \cref{fig:telegraph-issues}. 

\section{Leveraging Structured Data to Overcome Ambiguity}
\label{sect:struct}
These challenges highlight that snapshot counts alone are often insufficient for reliable mechanistic interpretation. \citet{jiao2024WhatCanWe} point toward a resolution: while steady-state distributions may be indistinguishable, transient dynamics often differ distinctly. Resolving these inherent ambiguities thus requires richer experimental data to provide additional constraints. Information embedded in temporal dynamics, spatial organization, or the relationships between multiple molecular species measured simultaneously holds the potential to distinguish between mechanisms that appear degenerate in static measurements. Leveraging these structured data types requires carefully constructing appropriate mechanistic models and robust methods to connect them with data and pinpoint what information can be extracted. Next, we will highlight recent case studies that exemplify successes in leveraging single-cell data structures for mechanistic insight.

\subsection{Temporal Dynamics Reveal Absolute Rates and Pathway Complexity}

Observing systems changing over time is the most intuitive way to resolve ambiguities inherent to static, snapshot-based inference, where often only ratios of kinetic rates can be inferred. Temporal data provides the necessary information to overcome this degeneracy and pinpoint absolute timescales. Such data in single-cell gene expression takes a variety of forms. The ideal dataset for any temporal inference is a genuine time series, where temporal correlations encode valuable information. While such systems exist (e.g., MS2/PP7 labeling \cite{tutucci2018ImprovedMS2System}) and have been used with mechanistic modeling \cite{burton2021InferringKineticParameters}, these measurements are challenging at scale. More commonly, temporal insights are gleaned from a series of destructive snapshots taken over time, where different cells are sampled, giving the marginal distributions of a population, as shown in \cref{fig:data-models}.  Even without connected time points, these temporal snapshots have seen recent success in distinguishing models and parameters, as discussed next.

Metabolic labeling techniques (e.g., using 4sU followed by sequencing protocols like scEU-seq or SLAM-seq  \cite{rodriques2021RNATimestampsIdentify, ramskold2024SinglecellNewRNA}) are a powerful tool to generate temporal measurements. These methods allow the distinction between newly synthesized ("new" or "labeled") and pre-existing ("old" or "unlabeled") RNA within single cells, often measured across time courses or after pulse-chase experiments. 
By tracking the dynamics of both labeled and unlabeled populations, one can decouple and directly estimate the absolute kinetic rates of RNA synthesis $\rho$ and degradation $\gamma$, parameters that are inherently intertwined from static measurements. \citet{volteras2024GlobalTranscriptionRegulation} demonstrate the power of this approach by integrating time-resolved metabolic labeling data to resolve absolute rates. Beyond the inference of rate parameters, the authors ask: how much temporal heterogeneity can be extracted from these measurements? Remarkably,  the authors can identify how the kinetic rates, e.g., $\rho(t), \gamma(t)$ vary temporally, averaging over technical complexities introduced by gene-doubling during replication. This genome-wide quantification of absolute, temporally-heterogeneous burst kinetics and mRNA degradation rates reveals complex regulatory patterns, such as distinct waves of burst frequency versus decay rate modulation across the cell cycle.

Beyond determining absolute rates, temporal data can also clarify the complexity of the underlying pathway, which we previously saw can be challenging with the telegraph model and snapshots (\cref{fig:telegraph-issues}, model mimicry). \citet{nicoll2025TransientPowerlawBehaviour} provides an analytical calculation that shows that for a large class of transcription models, the initial increase in mean mRNA levels following induction (assuming negligible initial counts and a known delay) follows a general power law, $\langle m(t) \rangle \propto t^n$. The exponent $n$ reflects the number of sequential, irreversible rate-limiting steps (promoter transitions, splicing, export) for mature transcript production. Measuring this transient scaling behavior, potentially in the minutes following induction, thus provides a method, robust to certain types of extrinsic noise, to estimate a lower bound on pathway complexity, distinguishing between models (e.g., with different numbers of inactive promoter states) that might yield identical steady-state distributions.

Surprisingly, even without any direct temporal measurements, modeling temporal variations is a valuable pursuit. For instance, \citet{sukys2025CellcycleDependenceBursty} derive a latent cell cycle phase $\tau$, computationally inferred from scRNA-seq snapshots, as a biological pseudo-time axis. By fitting mechanistic models that explicitly incorporate this latent cell-cycle progression, $\rho(\tau)$, they estimate phase-specific burst parameters for thousands of genes even without temporal measurements. Their findings reveal dosage compensation mechanisms (e.g., burst frequency halving post-replication while burst size remains constant) and highlight that neglecting such inherent temporal structure can significantly bias kinetic parameter estimates derived from naive snapshot analysis.
Popular RNA velocity methods also extract pseudotemporal information from snapshots by bridging the temporal component of this review with multimodal data (see \cref{subsect:multimodal}), particularly nascent and mature RNA. Recent variants employ fully mechanistic models to extract interpretable kinetics (e.g., \cite{gao2024DissectionIntegrationBursty, gayoso2024DeepGenerativeModeling,fang2025TrajectoryInferenceSinglecell}). 
These examples highlight how extracting temporal information is possible even without temporally connected time points. Understanding what information can (or cannot) be extracted from temporal snapshots (marginal distributions) remains an active area of fundamental theoretical research \cite{maddu2024InferringBiologicalProcesses,guan2025IdentifyingDriftDiffusion}.

\subsection{Spatial Context Connects Location to Kinetics and Function}

Beyond temporal observation, the spatial organization of molecules within cells
and tissues is increasingly accessible through high-resolution imaging, ranging
from single-molecule FISH (smFISH) for quantifying a few specific gene
transcripts to highly multiplexed methods like MERFISH capable of mapping
hundreds simultaneously \cite{chen2015SpatiallyResolvedHighly}. These spatial patterns offer another valuable data modality to dissect mechanisms obfuscated in other measurements.
Various computational toolkits and statistical frameworks (e.g., \cite{wang2024ELLAModelingSubcellular}) have emerged to quantify mRNA enrichment in specific compartments (e.g., nucleus, cytoplasm, protrusions) or proximity to cellular structures. 
These methods offer functional insights, such as cell-type-specific pattern variations \cite{takei2025SpatialMultiomicsReveals}, but often lack mechanistic details of how spatial organizations arise.
Some mechanistic models (e.g., \cite{wang2025JointDistributionNuclear}) consider well-mixed compartments, overlooking the fine-grained spatial information available from imaging. 

Explicitly modeling molecular positions, however, unlocks the mechanistic link
between these fine-grained patterns and the underlying kinetics.
For instance, transcription occurs at a visible transcriptionally active site
(TAS), modeled as a point source at location $z$, within a nuclear domain $\Omega$ delineated by a fluorescent stain (e.g., DAPI). The subsequent approximate diffusive motion of mRNA through the crowded nucleoplasm (with diffusivity $D$), followed by processing, nuclear export, or degradation, encodes spatial mRNA patterns, although challenging to dissect due to the stochasticity and low molecular counts inherent to these systems. \citet{miles2025IncorporatingSpatialDiffusion} provides an example of quantifying these spatial gradients by coupling telegraph transcription kinetics with a spatial movement model and explicit nuclear export at a geometric boundary. While the resulting distribution of mRNA positions $x_1,\ldots, x_m$ is challenging to describe due to stochasticity in both the quantity and value, it is concisely encoded as a spatial Cox process \cite{schnoerr2016CoxProcessRepresentation}
\begin{subequations}
\begin{align}
x_1,\ldots, x_m &\sim \mathrm{Poiss}(u(x,T)),\\  \partial_t u = D\partial_{xx}u &- \gamma u + \rho \Lambda(t)\delta(x-z),\\
\kappa u(x,t) = -& D\partial_x u(x,t),\,\text{for } x\in\partial\Omega.
\end{align}
\end{subequations}
where $\kappa$ encodes nuclear export and $\Lambda(t)$ is the telegraph process that switches at rates $\alpha,\beta$. Crucially, this spatial information can resolve parameter ambiguities inherent in total counts. For example, if the diffusion coefficient $D$ is known (e.g., from live-cell tracking), the spatial patterns enable inference of production and degradation rates, which are otherwise confounded in non-spatial steady-state measurements, as we saw in \cref{sect:teleg}. Cell-to-cell heterogeneity in spatial features (e.g., nuclear shape, gene location) can also sometimes aid parameter identifiability 
\cite{miles2024InferringStochasticRates}. Altogether, harnessing spatial data offers unique opportunities to constrain transport kinetics and understand localization-dependent regulation.

\subsection{Multimodal Readouts Decouple Linked Processes}

\label{subsect:multimodal}

Measuring only a single molecular species, such as mature mRNA, necessarily integrates the effects of all upstream processes. This aggregation often makes it impossible to distinguish, for example, whether low mRNA levels are due to slow transcription or rapid degradation, see \cref{fig:telegraph-issues}.
Simultaneously measuring and modeling multiple linked molecular species provides intermediate readouts that can deconvolve connected processes, resolve ambiguities, and offer insight into unobservable aspects of the regulatory cascade \cite{ham2021PathwayDynamicsCan}.

A key example in transcriptomics is the joint measurement of nascent (unspliced) and mature (spliced) RNA, often achievable from standard scRNA-seq data \cite{sullivan2025AccurateQuantificationNascent}. The joint distribution of these counts $P(n,m)$ 
contains information on intermediate rates like splicing, which is lost when
only considering the marginal distribution of total mRNA.  
Fitting such multi-species models at scale, however, presents a significant computational challenge. To address this, \citet{carilli2024BiophysicalModelingVariational} developed biVI, which embeds a biophysical model within a variational autoencoder (VAE) framework. This hybrid VAE approach captures complex cell state heterogeneity in low dimensions while inferring interpretable, cell-state-dependent kinetic parameters from joint nascent/mature counts.

Further theoretical foundations for  multi-species models are rapidly advancing, from refined models of nascent/mature RNA dynamics \cite{gorin2022InterpretableTractableModels, gorin2022ModelingBurstyTranscription,szavits-nossan2023SteadystateDistributionsNascent} to compartment models that distinguish nuclear and cytoplasmic mRNA \cite{ma2024AnalysisDetailedMultistage,wang2025JointDistributionNuclear}. 
The latter have proven crucial for explaining observations of ultra-low, sub-Poissonian noise \cite{weidemann2023MinimalIntrinsicStochasticity} and are becoming practical for large-scale inference through new analytical solutions \cite{wang2025JointDistributionNuclear}. 
Together, these pursuits underscore the value of integrating tailored models with multimodal data.

\begin{table*}[htb]
\footnotesize %
\centering
% Adjust m{width} values as needed - using m{} for vertical centering
\begin{tabular}{@{}m{1.5cm} m{4.0cm} >{\raggedright\arraybackslash}m{7.5cm} m{1.5cm}@{}}
\toprule
\textbf{Methods} & \textbf{Core Principle} & \textbf{Tradeoffs (+ Strengths / - Limitations)} & \textbf{Examples} \\
\midrule
Likelihood-Based & Evaluate $P(\text{Data}|\text{Model, }\theta)$ using likelihood derived exactly (e.g., inverting GF) or approximately (e.g., FSP)
&   \begin{itemize}[nosep,leftmargin=*]
        \item[+] Strong statistical grounding
        \item[+] Enables standard UQ \& ID analysis (FIM, Profiles)
        \item[-] Requires tractable likelihood, limits model complexity
        \item[-] Can be computationally expensive
    \end{itemize}
&   {\tiny \cite{vo2023AnalysisDesignSinglecell, gorin2023StudyingStochasticSystems, kilic2023GeneExpressionModel,miles2025IncorporatingSpatialDiffusion,wang2025JointDistributionNuclear}}  \\
\midrule
Moment-Based & Match model's statistical moments (or other summaries) to data estimates
&   \begin{itemize}[nosep,leftmargin=*]
        \item[+] Computationally scalable, avoids full likelihood calculation
        \item[-] Loses distributional information, prone to bias
        \item[-] Sensitive to moment choice
        \item[-] UQ/ID challenging (bounds sometimes possible \cite{li2025MomentbasedParameterInference})
    \end{itemize}
& {\tiny \cite{grima2024QuantifyingCorrectingBias, nicoll2025TransientPowerlawBehaviour}}  \\
\midrule
Simulation-Based (SBI) & Compare data to simulations (via stats/dist/NNs). Uses e.g., ABC, NPE/NLE
&   \begin{itemize}[nosep,leftmargin=*]
    \item[+] Handles intractable likelihoods \& complex models
    \item[+] Neural: sim-efficient; learns features implicitly
    \item[+] ABC: simpler setup
    \item[-] Computationally intensive (esp. ABC)
    \item[-] Setup sensitive (ABC: stats; Neural: architecture/training)
    \item[-] Validation essential (approx. posteriors, reliability, ID)
        \end{itemize}

&  {\tiny \cite{tang2023ModellingCaptureEfficiency, volteras2024GlobalTranscriptionRegulation,sierra2024AIpoweredSimulationbasedInference}} \\
\midrule
Hybrid (ML-Enhanced) & Use ML (NNs, GNNs, VAEs) to approx./accelerate/embed model components
&   \begin{itemize}[nosep,leftmargin=*]
        \item[+]  Scalable and flexible; handles complex data/dynamics
        \item[+] Can embed mechanism for interpretability, e.g., biVI \cite{carilli2024BiophysicalModelingVariational}
        \item[-]  Surrogate errors can bias inference; "black box" risks
        \item[-] Challenging UQ and validation; can obscure mechanistic ID
    \end{itemize}
& {\tiny \cite{jiang2021NeuralNetworkAided,carilli2024BiophysicalModelingVariational, cao2024EfficientScalablePrediction}}  \\
\bottomrule
\end{tabular}
\caption{Summary of inference techniques for stochastic models. UQ: uncertainty quantification, ID: identifiability, GF: generating functions, FSP: finite-state projection, NPE/NLE: neural posterior/likelihood estimation, ML: machine learning, NN: neural network, GNN: graph neural network, VAE: variational autoencoder, ABC: approximate Bayesian computation. \label{tab:inference_techniques}
 } 
%\vspace{0.5em}
\end{table*}

\section{Inference in Practice}

The previous section highlights how structured data and richer models can, in principle, provide deeper mechanistic insights. However, increased complexity introduces challenges in connecting structured data with models. For instance, how does one infer cell-cycle dependent forms like $\rho(t)$ rather than a simple rate $\rho$? Or how should one handle unwieldy multivariate distributions from multimodal or spatial data, especially with cellular heterogeneities? These newly introduced obstacles amplify the need for carefully chosen inference strategies that push the boundary of classical inference. 
This section reviews emerging inference strategies for stochastic models, highlighting tradeoffs, practical considerations, and successful applications in translating structured single-cell data into reliable biological insight, as summarized in \cref{tab:inference_techniques}.

\subsection{The Limits of Classical Inference for Complex Models}

Classical statistical inference hinges on evaluating the likelihood function, $L(\theta)=P(\text{data}|\text{model},\theta)$. For molecular counts, often $\log L(\theta) = \sum_{m} N(m)\log P(m|\theta)$, where $N(m)$ is the number of cells with RNA counts $m$. For simple models like the telegraph model, $P(m)$ can be computed analytically. For more complex models, this approach becomes challenging but can still be fruitful, particularly through two distinct directions.
The first is to compute $P(m)$ using generating functions, $G(z)=\sum_{m}z^m P(m)$, which can be derived with lengthy calculations for surprisingly complex models \cite{gorin2023StudyingStochasticSystems,wang2025JointDistributionNuclear} and then inverted numerically to recover $P(m)$. Second, Finite State Projection (FSP) methods, which truncate the state space $m=0,1,\ldots, M$ in a systematic way, provide approximate likelihoods for complex models \cite{vo2019BayesianEstimationStochastic, fox2019FiniteStateProjection,kilic2023GeneExpressionModel}. 
If full likelihood computation is infeasible but some analytical progress is possible, moment-based methods match model-derived statistical moments (e.g., time-dependent means \cite{nicoll2025TransientPowerlawBehaviour}) to data estimates, circumventing likelihood calculation. Moment-based approaches inherently lose distributional information, and uncertainty quantification can be challenging, although recent work shows promise in deriving parameter bounds \cite{li2025MomentbasedParameterInference}. At the frontier of addressing the computational burden of computing these quantities for complex models, efficient approximation methods like Holimap \cite{jia2024HolimapAccurateEfficient} aim to map complex network dynamics onto simpler, solvable systems. 

\subsection{Expanding the Toolkit: Simulation-Based and ML-Enhanced Inference}

When likelihood or moment calculations are infeasible for models with complex data structures, simulation-based inference (SBI) offers a practical alternative that uses forward simulations to infer parameters without an explicit likelihood. The most classical SBI approach, approximate Bayesian computation (ABC), compares simulations with data through the lens of user-chosen summary statistics.
Parameters with simulations ``close'' to the data are accepted to build an approximate posterior distribution. This compression into lower-dimensional statistics enables data-model comparison to be feasible, but the performance of ABC critically depends on the choice of context-specific summary statistics that are sufficiently informative.  
For example, to infer absolute rates and cell-cycle dependence from complex metabolic labeling data, \citet{volteras2024GlobalTranscriptionRegulation} use dynamic correlations between labeled/unlabeled RNA and moments (means, variances) of the populations to pin down dynamics and how they vary temporally. 

While conceptually straightforward, ABC's critical dependence on manually chosen summary statistics motivates the use of modern alternatives. Neural SBI approaches (e.g., neural posterior/likelihood estimation, NPE/NLE) employ deep learning to train neural networks on simulation outputs as surrogate approximations to the posterior distribution  $P(\theta | \text{data})$ or the likelihood function. In doing so, neural SBI methods implicitly learn informative features and show promise for superior computational efficiency over ABC \cite{lueckmann2021BenchmarkingSimulationBasedInference,frazier2024StatisticalAccuracyNeural,ramirezsierra2025ComparingAIOptimization}. However, enjoying these benefits is not entirely riskless. 
The challenge shifts to decisions in network architecture, training procedures, and the crucial step of validating the surrogate. Neural SBI methods can yield overconfident posteriors when the underlying model is misspecified \cite{kelly2024MisspecificationrobustSequentialNeural} and is complicated by the difficulty of validating the surrogate's accuracy. \cite{ramirezsierra2025ComparingAIOptimization}. 

There is growing momentum in combining deep learning's representational power with the mechanistic interpretability of classical approaches. Hybrid methods at this intersection have emerged that leverage neural networks to accelerate simulations, create surrogates for model components, or embed mechanistic parameters within flexible representations. For instance, biVI \cite{carilli2024BiophysicalModelingVariational} uses a variational autoencoder (VAE) with analytically-computed basis functions that describe an otherwise intractable joint distribution of nascent and mature mRNA counts. \citet{sukys2025CellcycleDependenceBursty} demonstrate another clever hybrid by using deep learning to extract a latent cell-cycle time $\tau$ for each observation, which is then used in a more traditional moment-approximated inference scheme to reveal how kinetic rates vary across the cell cycle, e.g., $\rho(\tau)$. These two examples highlight the merit of hybrid methods, where careful construction of the interface between mechanistic and machine learning components can yield powerful interpretability, computational tractability, and statistical expressiveness.

\subsection{Tradeoffs and Reliability}

Beyond models of gene expression, deep-learning-based methods for general stochastic reaction models \cite{jiang2021NeuralNetworkAided,huang2024DeepLearningLinking} and spatial stochastic models  \cite{ramirezsierra2025ComparingAIOptimization} show remarkable promise for scalability and flexibility. For instance, graph neural networks trained on simple geometries can extrapolate predictions of stochastic spatial reaction models to new complex geometries \cite{cao2024EfficientScalablePrediction}. 

Despite their promise, these deep-learning methods must be used with caution and undergo extensive validation to avoid the ``black box'' issues that can confound their output  \cite{ahlmann-eltze2025DeepLearningbasedPredictions}. This need for scrutiny arises because such methods often rely on neural network surrogates to approximate the true posterior or likelihood, and the reliability of the final inference hinges on the fidelity of this approximation \cite{ramirezsierra2025ComparingAIOptimization}. The choice of what these surrogates approximate, e.g., the posterior or the likelihood, determines whether the valuable toolkit of classical inference is forfeited or made approximate. Methods that directly learn an approximate posterior, such as Neural Posterior Estimation (NPE), largely bypass the likelihood and its associated toolkit. In contrast, other strategies, including the hybrid approaches discussed previously, use neural networks to creatively compute an \textit{approximate} likelihood. These methods can serve as standalone surrogates for the likelihood function itself (i.e., Neural Likelihood Estimation, NLE) or be embedded within larger models to resolve specific bottlenecks, such as approximating an intractable distribution \cite{carilli2024BiophysicalModelingVariational} or learning unknown terms to render a master equation solvable \cite{cao2024EfficientScalablePrediction}. In these likelihood-approximating cases, the classical toolkit remains accessible in principle, but its power for evaluating obstacles like practical identifiability and model mimicry is now contingent on the surrogate's accuracy. This toolkit spans profile likelihood analysis to evaluate parameter unidentifiability \cite{simpson2023ProfilewiseAnalysisProfile}, likelihood-based information criteria for model selection \cite{kilic2023GeneExpressionModel}, and information-theoretic quantities (e.g., the Fisher information) to direct optimal experimental design \cite{vo2023AnalysisDesignSinglecell,cook2024SequentialDesignSinglecell}. By comparison, the equivalent toolkit for modern, deep-learning-based inference is under active development but remains less refined, marking an active frontier for the field.

\section{Conclusion and Outlook}

Recent advances in mechanistic inference for stochastic gene expression signal a maturing field, moving beyond simply fitting models to critically evaluating what can be reliably learned from available data.  The increasing availability of structured single-cell datasets (whether temporal, spatial, or multimodal) has created new opportunities to resolve regulatory mechanisms, while also sharpening the need for rigorous validation and principled model selection. The next frontier lies in integrating an even richer array of molecular measurements. This expansion, however, introduces formidable new challenges alongside immense promise. 

Experimental techniques are pushing the field along several, often intertwined, axes of complexity. One axis pushes toward greater molecular detail for single genes. For instance,  long-read sequencing \cite{belchikov2024UnderstandingIsoformExpression} and imaging \cite{dingConstitutiveSplicingEconomies2019} methods can distinguish mRNA isoforms  or CRISPR-based live-cell RNA tracking \cite{xia2025SinglemoleculeLivecellRNA}  encode post-transcriptional dynamics with unprecedented detail. Another axis expands the breadth of measurement. Genuinely multi-omic datasets that pair transcriptomics with epigenomics (e.g., chromatin accessibility) or proteomics (e.g., surface protein levels) are now widely available. These modalities hold tremendous potential for creating more complete regulatory models but require synthesizing data types that differ fundamentally in their biological timescales and technical biases \cite{baiao2025TechnicalReviewMultiomics}. A third axis scales these models upward to larger systems like gene regulatory networks (GRNs) and multicellular tissues \cite{joly-smith2021InferringGeneRegulation,herbach2023HarissaStochasticSimulation}. Applying these insights at such scales magnifies inferential challenges due to a combinatorial explosion of potential interactions, making it difficult to infer causal regulatory links from transcriptomic data alone \cite{kernfeld2024TranscriptomeDataAre}.

Tackling these frontiers underscores this review's central thesis: the principled approaches developed for single-gene inference are not merely an inspiration but constitute the essential groundwork for this next era. They provide the mechanistic building blocks for larger models and, critically, a conceptual framework for disentangling the contributions of intrinsic, extrinsic, and technical noise, a task that becomes paramount in more complex systems. Ultimately, as experimental techniques advance, the primary bottleneck shifts from data acquisition to robust, interpretable analysis. Extracting reliable insights requires more than rich data; it demands an inference ecosystem that integrates experimental design with rigorous diagnostics.  While tools for identifiability analysis, sensitivity diagnostics, and principled model comparison are well-developed alongside classical inference, a critical gap remains in adapting them for the simulation-based and deep-learning methods essential for these complex systems \cite{noordijk2024RiseScientificMachine, ramirezsierra2025ComparingAIOptimization}. Building generalizable, modular tools that integrate these diagnostics directly into the inference workflow will be the crucial step in ensuring our conclusions are truly meaningful.

\section*{Acknowledgements}

The work was partially supported by NSF CAREER DMS-2339241 and NSF/NIGMS DMS-2451263. The author thanks Fangyuan Ding, Alex Mogilner, and Elizabeth Read for helpful comments on an earlier draft, and the anonymous reviewers for their careful reading and thoughtful suggestions.

\section*{Generative AI Declaration}

During the preparation of this work, the author used Gemini 2.5 Pro (Google) to improve language and readability. After using this tool/service, the author reviewed and edited the content as needed and takes full responsibility for the content of the publication.

\nomenclature{UQ}{Uncertainty Quantification}
\nomenclature{ID}{Identifiability}
\nomenclature{FSP}{Finite State Projection}
\nomenclature{FIM}{Fisher Information Matrix}
\nomenclature{MLE}{Maximum Likelihood Estimation}
\nomenclature{ABC}{Approximate Bayesian Computation}
\nomenclature{NPE}{Neural Posterior Estimation} 
\nomenclature{NLE}{Neural Likelihood Estimation}
\nomenclature{NN}{Neural Network}
\nomenclature{GNN}{Graph Neural Network}
\nomenclature{VAE}{Variational Autoencoder}
\nomenclature{MGF}{Moment Generating Function} % Added MGF

%\printnomenclature

{\scriptsize
\section*{Annotated Key Citations}

•• \citet{volteras2024GlobalTranscriptionRegulation}  Quantifies genome-wide, cell-cycle-dependent burst kinetics and degradation rates using time-resolved metabolic labeling, stochastic modeling, and scalable ABC inference.

•• \citet{jiao2024WhatCanWe} Demonstrates that fitting the simple telegraph model to steady-state data from complex gene expression processes can yield misleading parameters, highlighting model mimicry.

•• \citet{carilli2024BiophysicalModelingVariational} Introduces biVI, a hybrid VAE framework embedding biophysical models to infer transcriptional kinetics from multimodal nascent/mature scRNA-seq data

••\citet{nicoll2025TransientPowerlawBehaviour} Shows theoretically that transient power-law scaling of mean mRNA following induction distinguishes mechanistic models indistinguishable at steady state.

•\citet{sukys2025CellcycleDependenceBursty} Estimates cell-cycle phase-dependent burst parameters by fitting mechanistic models that account for extrinsic noise, using deep learning to infer cell age from scRNA-seq. 

•\citet{ramirezsierra2025ComparingAIOptimization}  Compares neural SBI (SNPE) with simulated annealing for parameter inference of spatial-stochastic models, assessing posterior quality and computational tradeoffs.

%Vo et al. (2023) - Analysis and Design... Rejecting Measurement Noise (*) Annotation: Extends Fisher Information Matrix-based optimal experimental design to account for measurement noise in single-cell experiments.

(\textit{•• = of outstanding interest; • = of special interest})
}

{\scriptsize 
  \bibliographystyle{elsarticle-num-names} 
\bibliography{inference_review}
}
%\bibliography{preprint}

\end{document}